%% file: Market_Stability.tex
\documentclass[letterpaper,10pt,conference]{ieeeconf}
\pdfoutput=1

\usepackage{color}
\usepackage{times}
\usepackage[dvipsnames]{xcolor}
\usepackage{amsmath, mathtools}
\usepackage{amssymb}
\usepackage{graphicx}
\usepackage{epsfig}
\usepackage{bm}
\usepackage{url}
\usepackage[font=footnotesize, skip=4pt]{caption}
\usepackage{subfig}
\usepackage{amsthm}

\newtheorem{theorem}{Theorem}

\newtheorem{assumption}{Assumption}

\IEEEoverridecommandlockouts
\usepackage{tikz}
\usetikzlibrary{positioning}
\tikzstyle{status} = [rectangle, draw=black, text centered, anchor=north, text=black, minimum width=15em, minimum height=4em, node distance=6ex and 7em]
\tikzstyle{line} = [draw,thick,-latex]
\tikzstyle{transition} = [font=\small]

\input{defs.tex}

\begin{document}


\title{\LARGE \bf Stability Analysis of Wholesale Electricity Markets under\\Dynamic Consumption Models and Real-Time Pricing}
\author{Datong P. Zhou$^{\ast\dagger}$, Mardavij Roozbehani$^\dagger$, Munther A. Dahleh$^\dagger$, and Claire J. Tomlin$^\star$
\thanks{$^\ast$Department of Mechanical Engineering, University of California, Berkeley, USA. {\tt\footnotesize datong.zhou@berkeley.edu}}
\thanks{$^\dagger$Laboratory for Information and Decision Systems, MIT, Cambridge, USA.
{\tt\footnotesize [datong,mardavij,dahleh]@mit.edu}}
\thanks{$^\star$Department of Electrical Engineering and Computer Sciences, University of California, Berkeley, USA.
{\tt\footnotesize tomlin@eecs.berkeley.edu}}%
\thanks{This work has been supported in part by the National Science Foundation under CPS:FORCES (CNS-1239166).}
}

\maketitle
\thispagestyle{empty}
\pagestyle{empty}

\begin{abstract}
This paper analyzes stability conditions for wholesale electricity markets under real-time retail pricing and realistic consumption models with memory, which explicitly take into account previous electricity prices and consumption levels. By passing on the current retail price of electricity from supplier to consumer and feeding the observed consumption back to the supplier, a closed-loop dynamical system for electricity prices and consumption arises whose stability is to be investigated. Under mild assumptions on the generation cost of electricity and consumers' backlog disutility functions, we show that, for consumer models with price memory only, market stability is achieved if the ratio between the consumers' marginal backlog disutility and the suppliers' marginal cost of supply remains below a fixed threshold. Further, consumer models with price and consumption memory can result in greater stability regions and faster convergence to the equilibrium compared to models with price memory alone, if consumption deviations from nominal demand are adequately penalized.
\end{abstract}


\input{Sections/Introduction}

\input{Sections/Market_Model}

\input{Sections/Consumption_Models}

\input{Sections/Stability_Analysis}
\input{Sections/Conclusion}


\bibliographystyle{IEEEtran}
\bibliography{bibliography}

\input{Sections/Appendix}

\end{document}

%% file: defs.tex
\makeatletter
\newcommand{\revisionhistory}[1]{%
\@ifundefined{showrevisionhistory}{\relax}{%
{#1}%
}%
}
\makeatother

%% file: Sections/Introduction.tex
%
%


\section{Introduction}
\label{sec:Introduction}
With the ever-growing demand for electricity, increased penetration of inherently volatile renewable energy resources, and the lack of economical energy storage technologies, wholesale prices of electricity fluctuate by up to an order of magnitude during peak and low demand periods. From an economic perspective, researchers have long been arguing in favor of real-time electricity pricing to remedy allocative inefficiencies associated with classic, near-constant energy tariffs such as Time-of-Use Pricing, see e.g. \cite{Borenstein:2005aa, Allcott:2011aa}. In \cite{Borenstein:2002aa}, the authors state practical implications of real-time pricing such as load flattening and peak shaving. Despite improved market efficiency of real-time pricing, ratepayers are hesitant towards their adoption due to possible sudden price spikes \cite{Borenstein:2005ab}. While we acknowledge the risk aversion of ratepayers, we neither argue in favor nor against real-time pricing, but restrict our attention to the analysis of conditions under which real-time pricing results in stable price evolutions.

The extant body of literature on the dynamics of electricity markets is rich and multifaceted. A common approach is load scheduling in a Demand Response setting with time-varying, exogenous pricing, where the elastic component of demand, i.e. the shiftable part, is to be optimally satisfied to minimize cost or, equivalently, maximize utility \cite{Mohsenian-Rad:2010aa}, \cite{Conejo:2010aa}, \cite{Chen:2010aa}. Optimal utilization policies and efficiency gains with storage are investigated in \cite{Faghih:2011aa, Bitar:2011aa}. \cite{Huang:2015aa} has examined a tradeoff between risk and efficiency under cooperative and non-cooperative scheduling behavior. These contributions rely on strong parameterizations of household appliances or agent arrival processes.

In this paper, we abstract away those assumptions, and instead model the real-time wholesale electricity market dynamics as a closed-loop feedback system between the suppliers' electricity generation and the consumers' electricity consumption, which arises by providing end-users of electricity with a proxy of the current electricity retail price. Such systems have been studied and analyzed for stability in the extant literature, see for example \cite{Roozbehani:2011aa} where the authors investigate theoretical statements on the utility functions under which the market remains stable, and \cite{Alvarado:1999aa}, which models market dynamics with a system of differential algebraic equations. However, in these works, the suppliers' and consumers' decision on the quantity to generate and consume, respectively, are explained with \textit{invariant} cost functions which remain constant over time. While this is a tenable assumption for the suppliers' cost, a \textit{dynamic} user consumption model which incorporates previous consumption and electricity prices into the consumption decision is more suitable to capture time-varying user preferences. Proposing these dynamic consumption models with memory is a key contribution of this paper.

More specifically, we formulate the optimal consumption strategy of suppliers as a constrained finite horizon control problem that satisfies the Bellman Equation and is solvable with dynamic programming. At each time step, the user has the option to defer consumption, but thereby causing backlog associated with a disutility. In the derivation of these models, we assume the electricity prices to be exogenous and the suppliers to be price-taking. The stepwise optimal feedback policy is then identified as the new consumption model with memory, as it depends on the current and previous electricity price as well as potentially the previous consumption. This consumption model is then used in a closed-loop setting under the assumption of endogenous prices (in particular, the supplier is not price taking anymore), to see whether or not this ``endogenous transformation" is stable. 

The key observation is that, under mild assumptions, this consumption model results in price and consumption stability if and only if the ratio between the consumers' marginal backlog disutility and the suppliers' marginal generation cost falls below a fixed threshold. This ratio can be increased by suitably penalizing consumption deviations from the required, a-priori known demand.

This paper is organized as follows: In Section \ref{sec:Electricity_Market_Model}, we describe the market setting of real-time pricing and the roles of the suppliers, consumers, and Independent System Operator, and set up a dynamical system for the temporal evolution of supply and demand. In Section \ref{sec:Consumption_Models}, we identify two memory-based consumption models. Next, Section \ref{sec:Stability_Analysis} investigates stability conditions under the memory-based consumption models within the dynamical system set up in Section \ref{sec:Electricity_Market_Model}. Section \ref{sec:Conclusion} concludes the paper.

%% file: Sections/Market_Model.tex
%
%


\section{Electricity Market Model}
\label{sec:Electricity_Market_Model}
In this Section, we describe a model that describes the market participants, namely the energy suppliers, consumers, and the independent system operator (ISO), and their interactions that result in a dynamical system whose stability is to be analyzed in Section \ref{sec:Stability_Analysis}.

\subsection{Market Participants}
\subsubsection{Suppliers}
Each supplier $i$ in the set of suppliers $\mathcal{S}$ is endowed with a cost function $c_i(\cdot):\mathbb{R}_+ \mapsto \mathbb{R}_+$, which maps a production quantity to its associated cost. The suppliers are price-taking and rational, profit-maximizing agents.
\begin{assumption}\label{as:cost_function_convex}
$c_i, i\in\mathcal{S}$ is convex and increasing.
\end{assumption}
Let $\lambda$ denote the unit price of electricity set by the ISO at which the suppliers are reimbursed. Then, with Assumption \ref{as:cost_function_convex}, supplier $i$'s production quantity $s_i$ is determined as
\begin{equation*}
s_i(\lambda) = \arg\max_{x\in\mathbb{R}_+} \lambda x - c_i(x) = \dot{c}_i^{-1}(\lambda)
\end{equation*}

\subsubsection{Consumers}
Each consumer $j$ in the set of consumers $\mathcal{D}$ is endowed with a utility function $v_j(\cdot):\mathbb{R} \mapsto \mathbb{R}$ and a time-varying, memory-based consumption model $u_j$, which is either a function of the current price $\lambda_k$, the previous price $\lambda_{k-1}$, and the previous consumption $u_{k-1}$ (Section \ref{sec:cons_model_price_cons_memory}), or just of the prices $\lambda_k$ and $\lambda_{k-1}$ (Section \ref{sec:cons_model_price_memory}). This is a deviation from common practice, as we explicitly include price and consumption memory into users' consumption policies, as opposed to static consumption models that determine the consumption as the inverse of the derivative of the utility function evaluated at the current price (e.g. \cite{Roozbehani:2014aa, Li:2011aa}).

\subsubsection{Independent System Operator}
The ISO's task is to solve the economic dispatch problem \cite{Wu:1996aa}, i.e. to ensure supply follows demand at all times while minimizing transmission cost subject to transmission, capacity, and congestion constraints. For ease of exposition, we assume zero resistive losses, infinite capacity limits, and the absence of congestion in this paper. For an analysis of market operation under these constraints, the reader is referred to \cite{Gomez-Exposito:2016aa}.

Under the above assumptions, the ISO's optimization problem at each time step can be formulated as
\begin{equation}\label{eq:ISO_optimization_problem}
\begin{aligned}
& \underset{\lbrace u_j\rbrace_{j\in\mathcal{D}},\lbrace s_i\rbrace_{i\in\mathcal{S}}}{\text{maximize}}
& & \sum_{j\in \mathcal{D}} v_j(u_j) - \sum_{i \in \mathcal{S}} c_i(s_i)\\
& \text{subject to}
& & \sum_{j\in \mathcal{D}} u_j = \sum_{i\in \mathcal{S}} s_i
\end{aligned}
\end{equation}
where we have dropped time indices for notational ease. We emphasize that \eqref{eq:ISO_optimization_problem} has to be solved at each time step.

\eqref{eq:ISO_optimization_problem} assumes that consumers announce their value functions $\lbrace v_j\rbrace_{j\in\mathcal{D}}$ to the real-time wholesale market. However, this is an unrealistic assumption particularly for small consumers. Instead, we analyze the market setting in which the value functions are unknown (similar to \cite{Roozbehani:2012aa}). Further, we adopt the \textit{Representative Agent Model} \cite{Canova:2007aa} to reduce the set of consumers $\mathcal{D}$ and suppliers $\mathcal{S}$ to a singleton each, thereby implicitly assuming that all consumers and suppliers act such that the sum of their choices is equivalent to the individuals' decisions. This simplification allows for an analysis of the aggregate supply and demand dynamics rather than their microscopic evolutions. Let $c$ and $v$ denote the representative cost and value functions, respectively, $u$ the aggregate demand, and $s$ the aggregate supply. Lastly, since the set of value functions $\lbrace v_j\rbrace_{j\in\mathcal{D}}$ is now unknown to the ISO, the aggregate demand $\hat{u}$ has to be predicted by the ISO, which is assumed to be constant over each time interval. With the above-mentioned assumptions, \eqref{eq:ISO_optimization_problem} reduces to
\begin{equation}
\underset{s}{\text{min}}~ c(s)\quad\text{s.t.}\quad\hat{u}=s
\end{equation}
whose solution is simply $c(\hat{u})$ with associated unit price of electricity $\lambda = \dot{c}(\hat{u})$.

\subsection{Real-Time Supply-Demand Model}
\begin{figure}
\centering
\begin{tikzpicture}
    \node [status, align=left, fill={rgb:black,0.5;white,3}] (T) {$\hat{u}_{k+1} = u_k$ \\ $\lambda_{k+1} = \dot{c}(\hat{u}_{k+1})$};
    \node [status, fill={rgb:black,0.5;white,3}, below=2em of T] (A) {$u_{k+1} = \begin{cases}
      \eqref{eq:price_dynamics_no_memory}, & \mathrm{static} \\
      \eqref{eq:opt_cons_pric_memo_only}, & \mathrm{price~memory} \\
      \eqref{eq:cons_opt_flattening}, & \mathrm{price+cons.~memory}
    \end{cases}$};
	
	\path [line] (T) -- ++(4,0) -- +(0,-2.3) -- (A) node [transition,pos=0.0,right] {\large{$\lambda$}};
	\path [line] (A) -- ++(-4,0) -- +(0,2.3) -- (T) node [transition,pos=0.0,left] {\large{$u$}};
\end{tikzpicture}
\vspace{0.04cm}
\caption{Closed-Loop Feedback System Between Suppliers and Consumers}
\label{fig:Closed_Loop_Feedback_System}
\end{figure}
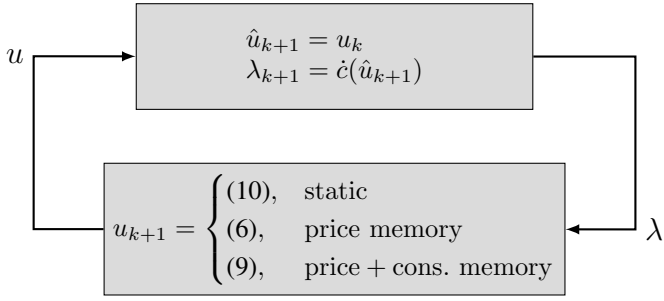

Figure \ref{fig:Closed_Loop_Feedback_System} depicts the closed loop dynamical system that determines the temporal evolution of supply and demand, coordinated by the price, which emerges under an ex-ante pricing system. Under ex-ante pricing, the ISO at time $k$ determines the wholesale price \textit{before} the next time $k+1$ based on the predicted consumption, $\hat{u}_{k+1}$. In this scenario, the gap between the predicted consumption $\hat{u}_{k+1}$ and the actual consumption $u_{k+1}$ results in a price difference between the ex-ante price $\lambda_{k+1}$, based on which the consumer demands electricity, and the actual price that materializes only after the consumption $u_{k+1}$ does, which is the price the supplier is reimbursed at. This gap could be either positive or negative, and the risk associated with it is assumed by the ISO. In \cite{Roozbehani:2011aa}, the authors show that an ex-post pricing system, where the wholesale price $\lambda_{k+1}$ is determined only after the consumption $u_{k+1}$ materializes, leads to identical price and consumption dynamics, but unlike ex-ante pricing, the consumer has to bear price uncertainty.

The electricity price $\lambda$ is now an endogenous process as it depends on the observed consumption. At time $k$, $\lambda_k$ is declared to the consumer and elicits a myopic adjustment of consumption based on her consumption model, which we assume to be time-varying and with memory (\eqref{eq:opt_cons_pric_memo_only} or \eqref{eq:cons_opt_flattening}). For completeness, a static utility function could be used, and the interested reader is referred to \cite{Roozbehani:2011aa, Roozbehani:2012aa} for a theoretical treatment of that case.

%% file: Sections/Consumption_Models.tex
%
%


\section{Electricity Consumption Models}
\label{sec:Consumption_Models}
In this Section, we derive consumption models with memory by casting the energy consumption as an inventory problem \cite{Zipkin:2000aa}. Suppose the consumer seeks the optimal policy to minimize cost over a given horizon with $n$ slotted intervals, which are indexed as $k = 0, \ldots, n-1$. Denote the price per unit of electricity at time $k$ as $\lambda_k$, which is assumed to be an exogenous, random process. At time $k$, $\lambda_k$ is announced to the consumer. The electricity demand at interval $k$ is denoted by $d_k$, assumed to be known in advance, and shiftable (elastic) such that it can be satisfied in the periods $[k, \ldots, n-1]$. The possibility to shift demand creates a backlog $x_k \leq 0$, which denotes the amount of unsatisfied energy at interval $k$. Further, let $u_k \geq 0$ denote the consumption withdrawn from the grid at interval $k$, and assume there is no storage, but the option to sell back energy to the grid at price $\lambda_k$. With the terminal constraint $x_n = 0$, the solution to this inventory problem is given as the minimizer of \eqref{eq:optimization_scheduling}.

\begin{align}
& \underset{u_0, \ldots, u_{n-1}}{\text{minimize}}
& & \mathbb{E}_{\lambda_1, \ldots, \lambda_{n-1}}\left[ \sum_{k=0}^{n-1} \lambda_k u_k + p(x_{k+1}) + h(u_k, d_k) \right] \nonumber \\
& \text{subject to}
& & x_{k+1} = x_k + u_k - d_k \label{eq:optimization_scheduling}\\
& & & x_k \leq 0 \nonumber \\
& & & x_n = 0 \nonumber
\end{align}
In \eqref{eq:optimization_scheduling}, $h(\cdot, \cdot):\mathbb{R}_+ \times \mathbb{R}_+ \mapsto \mathbb{R}_+$ is a cost term that penalizes deviations of the actual consumption $u_k$ from the demand $d_k$, and $p(\cdot):\mathbb{R}_- \mapsto \mathbb{R}_+$ the backlog penalty function. We make the following
\begin{assumption}\label{as:h_d_strictly_convex}
$h$ and $p$ are strictly convex in their first argument.
\end{assumption}

We note that the cost function in \eqref{eq:optimization_scheduling} satisfies the Bellman equation:
\begin{equation}
\begin{aligned}\label{eq:bellman_equation_cost}
J_k^\ast = \underset{u_k}{\text{min}}\quad &\lambda_k u_k + p(x_{k+1}) + h(u_k, d_k) \\
&+\mathbb{E}_{\lambda_{k+1}, \ldots, \lambda_{n-1}}\left[ J_{k+1}^\ast \right].
\end{aligned}
\end{equation}
$J_k^\ast$ denotes the $k$-step optimal cost-to-go. This allows us to derive the optimal consumption decision $u_k^\ast$ at any given time $k$ to construct consumption models which contain memory about past prices and decisions. In Section \ref{sec:cons_model_price_memory}, we explicitly solve \eqref{eq:bellman_equation_cost} for the case $h \equiv 0$ to obtain the optimal consumption $u^\ast$, an approximation of which is used in a subsequent stability analysis. A closed-form solution to \eqref{eq:bellman_equation_cost} for $h \neq 0$ does not exist, and so we express the consumption model in terms of the value function in Section \ref{sec:cons_model_price_cons_memory}. For a solution with \textit{linear}, time-specific backlog penalties, $h(u_k, d_k) \equiv 0$, and no sell-back option to the grid (i.e. $u_k\geq 0$), the reader is referred to \cite{Roozbehani:2014aa} and the references therein.

\subsection{Consumption Model with Price Memory}\label{sec:cons_model_price_memory}
First, we consider the case $h \equiv 0$ in \eqref{eq:optimization_scheduling}. We show that this leads to a consumption model with price memory, but no consumption memory. The optimization problem at any given time $k$ is formulated as

\begin{equation*}
\begin{aligned}
J_k^\ast =~ & \underset{u_k}{\text{min}}
& & \lambda_k u_k + p(x_{k+1}) + \mathbb{E}_{\lambda_{k+1}, \ldots, \lambda_{n-1}}\left[J_{k+1}^\ast\right] \\
& \text{s.t.}
& & x_{k+1} = x_k + u_k - d_k
\end{aligned}
\end{equation*}
\begin{assumption}\label{as:expected_price}
At any given time $k$, the consumer assumes 
$$\mathbb{E}\left[\lambda_{k+1}\right] = \lambda_k, \quad k=0, \ldots, n-2. $$
\end{assumption}

\begin{theorem}\label{thm:optimal_consumption_bellmann}
With Assumption \ref{as:expected_price}, the optimal consumption and cost-to-go are given by
\begin{subequations}
\begin{align}
u_{n-k}^\ast &= d_{n-k} - \dot{p}^{-1}\left( \frac{\lambda_{n-k}-\lambda_{n-k-1}}{k-1}  \right) \label{eq:opt_consumption_bmann}\\
J_{n-k}^\ast &= (k-1)\cdot p\left( x_{n-k} - \dot{p}^{-1}\left( \frac{\lambda_{n-k}-\lambda_{n-k-1}}{k-1} \right) \right)\nonumber \\
&\hspace{0.5cm}+\lambda_{n-k}\left( \sum_{i=n-k}^{n-1} d_i - x_{n-k} \right) \label{eq:opt_costtogo_bmann}
\end{align}
\end{subequations}
for $k = 2, \ldots, n$ in \eqref{eq:opt_consumption_bmann} and $k = 1, \ldots, n$ in \eqref{eq:opt_costtogo_bmann}. For $k=1$, the optimal consumption is $u_{n-1}^\ast = d_{n-1} - x_{n-1}$.
\end{theorem}


Note that due to Assumption \ref{as:h_d_strictly_convex}, $\dot{p}(\cdot)$ and $\dot{p}^{-1}(\cdot)$ are well-defined. We see that $u_k^\ast$ is a function not only of the current price $\lambda_k$, but also of the previous one $\lambda_{k-1}$.

Due to the time-varying nature of the optimal consumption \eqref{eq:opt_consumption_bmann}, we make a simplification and approximate $u_{n-k}^\ast$ as a time-invariant function
\begin{equation}\label{eq:opt_cons_pric_memo_only}
u_{n-k}^\ast = d_{n-k} - \dot{p}^{-1}(\lambda_{n-k} - \lambda_{n-k-1})
\end{equation}

\subsection{Model with Price and Consumption Memory}\label{sec:cons_model_price_cons_memory}
In this Section, we model $h(u_k, d_k)$ as a cost term that penalizes the squared difference of their arguments:
\begin{equation}\label{eq:deviation_cost_term}
h(u_k, d_k) = \rho(u_k - d_k)^2
\end{equation}
where $\rho \geq 0$ is a scalar weight.
The $k$-step optimization at time $k$ is  
\begin{equation*}
\begin{aligned}
& \underset{u_k}{\text{min}}
& & \lambda_k u_k + p(x_{k+1}) + h(u_k, d_k) + \mathbb{E}_{\lambda_{k+1}, \ldots, \lambda_{n-1}}\left[ J_{k+1}^\ast \right] \\
& \text{s.t.}
& & x_{k+1} = x_k + u_k - d_k
\end{aligned}
\end{equation*}
Unlike the previous consumption model (Section \ref{sec:cons_model_price_memory}), closed-form expressions for the optimal consumption $u_k^\ast$ and cost-to-go $J_k^\ast$ do not exist in this case. We therefore approximate the optimal consumption by first defining $p(x_{k+1}) + J_{k+1}^\ast =: V_{k+1}(x_{k+1})$ and then solving the first order optimality condition with respect to $u_k^\ast$
\begin{align}\label{eq:optimality_condition_p_c_memory}
0 &= \lambda_k + 2\rho(u_k - d_k) + \frac{dV_{k+1}(x_{k+1})}{dx_{k+1}}\frac{dx_{k+1}}{du_k}
\end{align}
where we make the following
\begin{assumption}\label{as:dot_V_linear}
$V$ is a quadratic function, i.e. $\dot{V}^{-1}(x) = \tilde{V}(x)$ is linear.
\end{assumption}
Assumption \ref{as:dot_V_linear} can be justified with Assumption \ref{as:dot_p_linear} and the fact that $J_{k+1}^\ast$ can be approximated as a quadratic function. Then $V$ is the sum of two quadratic functions. With Assumption \ref{as:dot_V_linear}, \eqref{eq:optimality_condition_p_c_memory} becomes
\begin{align}
x_{k+1} = \tilde{V}(2\rho(d_k-u_k) - \lambda_k) = x_k + u_k - d_k \nonumber
\end{align}
Solving for $u_k$ and replacing $x_k$ yields
\begin{align}
u_k^\ast = \frac{d_k + \tilde{V}(\lambda_{k-1} - \lambda_k + 2\rho (d_k - d_{k-1} + u_{k-1})}{2\rho\tilde{V} + 1}\label{eq:cons_opt_flattening}
\end{align}


Thus, $u_k^\ast$ is a function of the current price $\lambda_k$, the previous price $\lambda_{k-1}$, and previous consumption $u_{k-1}$, and so the consumption model has price and consumption memory.

%% file: Sections/Stability_Analysis.tex
%
%


\section{Analysis of Market Stability}
\label{sec:Stability_Analysis}

\subsection{Stability Under Static Consumption Model}
We assume a static utility and cost function of the consumer and producer, respectively. Under ex-ante pricing, the price dynamics are
\begin{equation}\label{eq:price_dynamics_no_memory}
\lambda_{k+1} = \dot{c}(\hat{u}_{k+1}) = \dot{c}(\dot{v}^{-1}(\lambda_k))
\end{equation}
\eqref{eq:price_dynamics_no_memory} is a nonlinear difference equation, and a closed loop solution, in general, does not exist. Theoretical statements about the Lyapunov stability of this system have been made in \cite{Roozbehani:2011aa, Roozbehani:2012aa}.

\subsection{Stability Under Consumption Model with Price Memory}\label{sec:Consumption_Price_Memory}
Here we use consumption model \eqref{eq:opt_cons_pric_memo_only} to analyze market dynamics. The price dynamics are
\begin{equation}
\begin{aligned}
\lambda_{k+1} &= \dot{c}(\hat{u}_{k+1}) = \dot{c}(u_k) \\
&= \dot{c}\left( \dot{p}^{-1}(-\lambda_k) - \dot{p}^{-1}(-\lambda_{k-1}) + d_k \right). \label{eq:price_dynamics_memory}
\end{aligned}
\end{equation}

\eqref{eq:price_dynamics_memory} is nonlinear and thus cannot be solved explicitly except for special cases. One such case which allows to make quantitative statements without numerical simulations arises under the following assumptions:
\begin{assumption}\label{as:dot_p_linear}
$p(\cdot):\mathbb{R}_- \mapsto \mathbb{R}_+$ is a quadratic function.
\end{assumption} 
\begin{assumption}\label{as:c_linear_function}
$c(\cdot):\mathbb{R}_+ \mapsto \mathbb{R}_+$ is a quadratic function.
\end{assumption}
While Assumptions \ref{as:dot_p_linear} and \ref{as:c_linear_function} appear to be strong, they maintain generality in that they capture the natural fact of increasing marginal cost and constitute the simplest approximation to other nonlinear, convex cost functions. We can now write \eqref{eq:price_dynamics_memory} as follows:
\begin{equation}
\lambda_{k+1} = \left(\dot{c}\circ \dot{p}^{-1}\right)(-\lambda_k) - \left(\dot{c}\circ \dot{p}^{-1}\right)(-\lambda_{k-1}) + \dot{c}(d_k)
\end{equation}
Let $c(x) := \alpha x^2 \therefore \dot{c}(x) = 2\alpha x$, $p(x) := \beta x^2 \therefore \dot{p}^{-1}(x) = \frac{1}{2\beta}x,~\alpha, \beta \neq 0$, and $\varepsilon := \alpha/\beta >0$. Then the price and consumption dynamics are 
\begin{subequations}
\begin{align}
\lambda_{k} &= -\frac{\alpha}{\beta}\lambda_{k-1} + \frac{\alpha}{\beta}\lambda_{k-2} + 2\alpha d_k \label{eq:price_dynamics_memory_linear} \\
u_k &= \frac{1}{2\beta}(\lambda_{k-1} - \lambda_k) + d_k \label{eq:cons_dynamics_memory_linear}
\end{align}
\end{subequations}
Equation \eqref{eq:price_dynamics_memory_linear} is a linear, non-homogeneous difference equation in the price $\lambda$ which can be readily solved for selected $d_k$. We analyze two such cases - constant and sinusoidal demand - in the following.
\begin{theorem}[Stability under Consumption Model with Price Memory]\label{thm:stability_price_memory}
The price dynamics under a consumption model with price memory are stable if and only if $0 \leq \varepsilon < 1/2$.
\end{theorem}
Theorem \eqref{thm:stability_price_memory} translates to the fact that the price dynamics \eqref{eq:price_dynamics_memory_linear} are stable if and only if the marginal production cost of electricity is less than half the marginal backlog disutility.

\subsubsection{Constant Demand}\label{sec:Constant_Demand}
For the case of constant demand $d_k = d~\forall k$, and assuming that the price and consumption converge to fixed points, these can be determined directly from \eqref{eq:price_dynamics_memory_linear} and \eqref{eq:cons_dynamics_memory_linear}:
\begin{align}
(\lambda^\ast, u^\ast) &= (2\alpha d, d)\label{eq:fixed_point_lambda_and_u}
\end{align}

\begin{theorem}\label{thm:price_dyn_price_memory}
The price dynamics under the consumption model with price memory are
\begin{equation}\label{eq:price_dynamics_solution_price_memory}
\lambda_k = c_1 x_1^k + c_2 x_2^k + 2\alpha d.
\end{equation}
where the constants $c_1$ and $c_2$ are determined from initial conditions.
\end{theorem}

\begin{figure}[hbtp]
\centering
\includegraphics[scale=0.19]{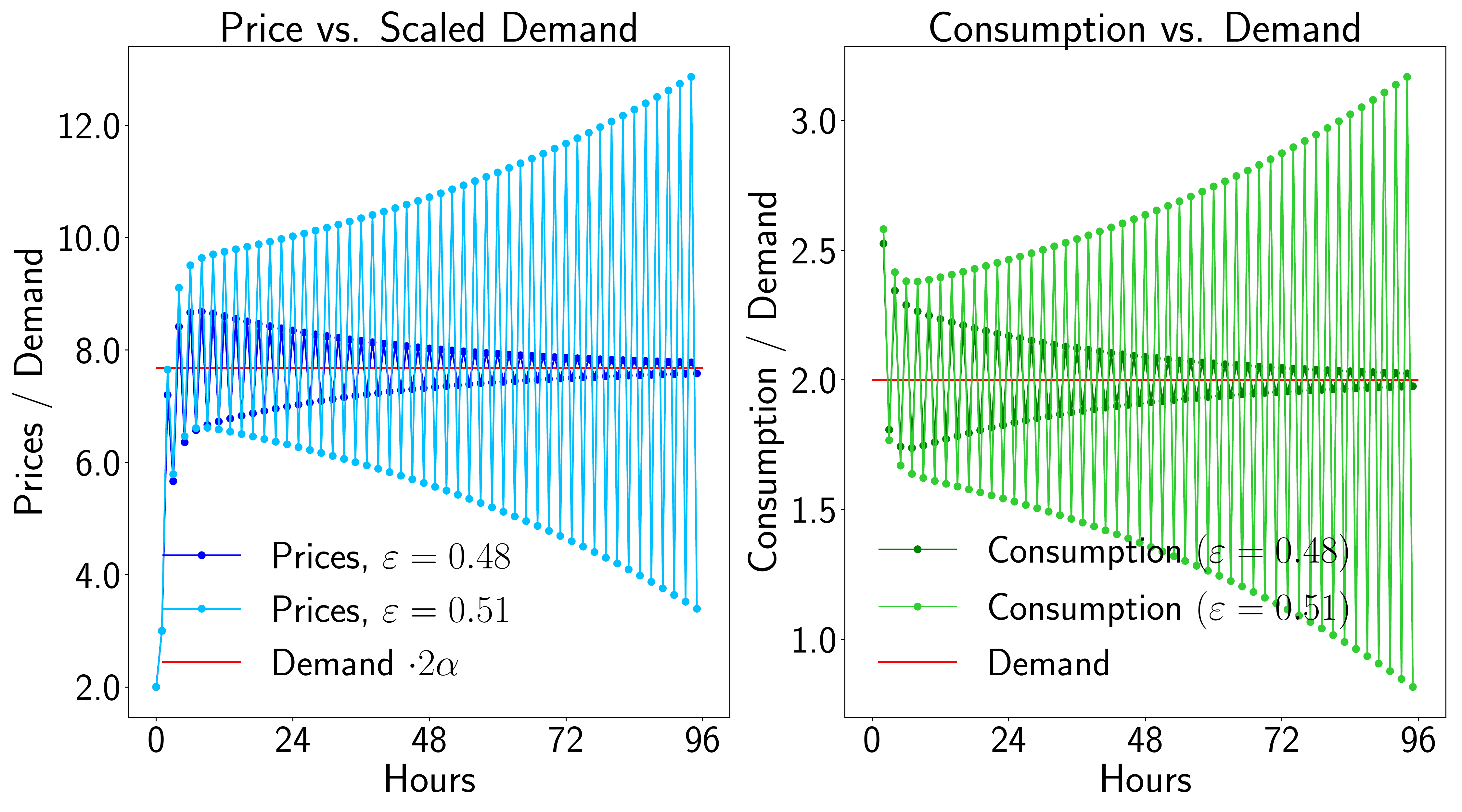}
\caption{Evolution of Prices and Consumption, Constant Demand, $\beta=4$, $\alpha = 1.92$ (Stable) or $\alpha = 2.04$ (Unstable). Initial Conditions $\lambda_0 = \lambda_1 = 3$}
\label{fig:Price_Recurrence_Const_Dem_Stable_Unstable}
\end{figure}

Figure \ref{fig:Price_Recurrence_Const_Dem_Stable_Unstable} shows the evolution of prices \eqref{eq:price_dynamics_memory_linear} and consumption for $\varepsilon = 0.48 < 0.5$, which is a stable system, and for $\varepsilon = 0.51 > 0.5$, which results in an unstable system.

\subsubsection{Variable Demand}\label{sec:Variable_Demand}
\begin{figure}[hbtp]
\centering
\vspace{-0.3cm}
\includegraphics[scale=0.19]{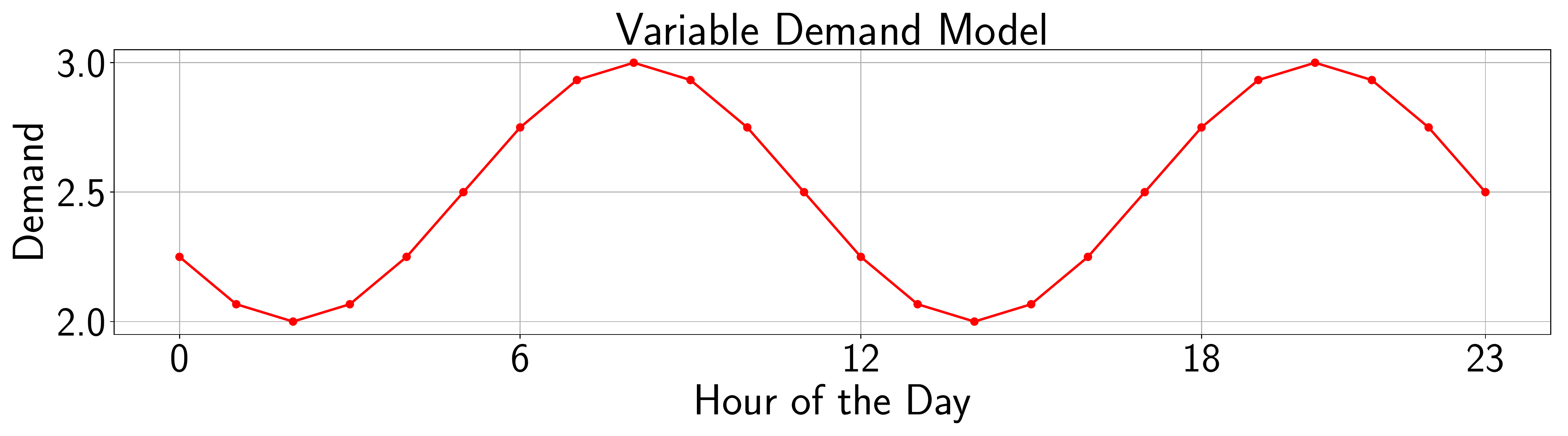}
\caption{Variable Demand Function Modeled as Sinusoid \eqref{eq:sinusoid_var_demand}}
\label{fig:sinusoid_var_demand}
\end{figure}

We model a variable demand curve as a sinusoid of period 12 hours, such that every day exhibits two full periods, see Figure \ref{fig:sinusoid_var_demand}.
Let
\begin{equation}\label{eq:sinusoid_var_demand}
2\alpha d_k = \mu + A\sin\left(\frac{(k-5)\pi}{6}\right)
\end{equation}
be the 12-hour periodic demand with mean $\mu$ and amplitude $A$, and let $k$ denote the hour of the day. As a consequence, the peak demand is attained at $k = 8$ or $k=20$, whereas the minima are at $k = 14$ and $k = 2$. This simplified load shape is an approximation of one of the most common residential load shapes with a double peak as identified in \cite{Zhou:2016aa}. Solving \eqref{eq:price_dynamics_memory_linear} and \eqref{eq:cons_dynamics_memory_linear} yields Theorem \ref{thm:price_dynamics_price_cons_memory}.

\begin{theorem}\label{thm:price_dynamics_price_cons_memory}
The price dynamics under the consumption model with price and consumption memory are
\begin{equation}
\begin{aligned}\label{eq:soln_rec_variable}
\lambda_k = &~c_1 x_1^k + c_2 x_2^k + \mu~+ \\
&e_1\sin\left(\frac{(k-5)\pi}{6}\right) + e_2\cos\left(\frac{(k-5)\pi}{6}\right).
\end{aligned}
\end{equation}
where the constants $c_1$ and $c_2$ are determined from initial conditions. $e_1$ and $e_2$ are
\begin{align*}
e_1 &= \frac{1 + \varepsilon(\sqrt{3}-1)/2}{1 + (\sqrt{3}-1)\varepsilon + (2 - \sqrt{3})\varepsilon^2}A \\
e_2 &= \frac{\varepsilon(1-\sqrt{3})/2}{1 + (\sqrt{3}-1)\varepsilon + (2 - \sqrt{3})\varepsilon^2}A 
\end{align*}
\end{theorem}

\begin{theorem}\label{thm:price_traj_limit}
For $\varepsilon = \alpha/\beta < 1/2$ and $k \rightarrow \infty$, the price trajectory converges to the limiting function
\end{theorem}
\begin{equation}\label{eq:price_traj_limit}
\begin{aligned}
\lambda_k \xrightarrow[]{k \to \infty}~ &\mu + \sqrt{e_1^2 + e_2^2}\cdot \sin \left( \frac{(k-5)\pi}{6}~+ \right.\\
&\left.\frac{\pi}{3}+\arctan\left( \frac{e_2 - \sqrt{3}e_1}{\sqrt{3}e_2 + e_1} \right) \right)
\end{aligned}
\end{equation}

Thus, as $\varepsilon$ increases, the magnitude of the resulting limiting sinusoid decreases, and we observe an additional negative phase shift, which translates into the fact that the prices as well as the consumption show a lagging behavior.

\begin{figure}[hbtp]
\centering
\vspace{-0.3cm}
\includegraphics[scale=0.19]{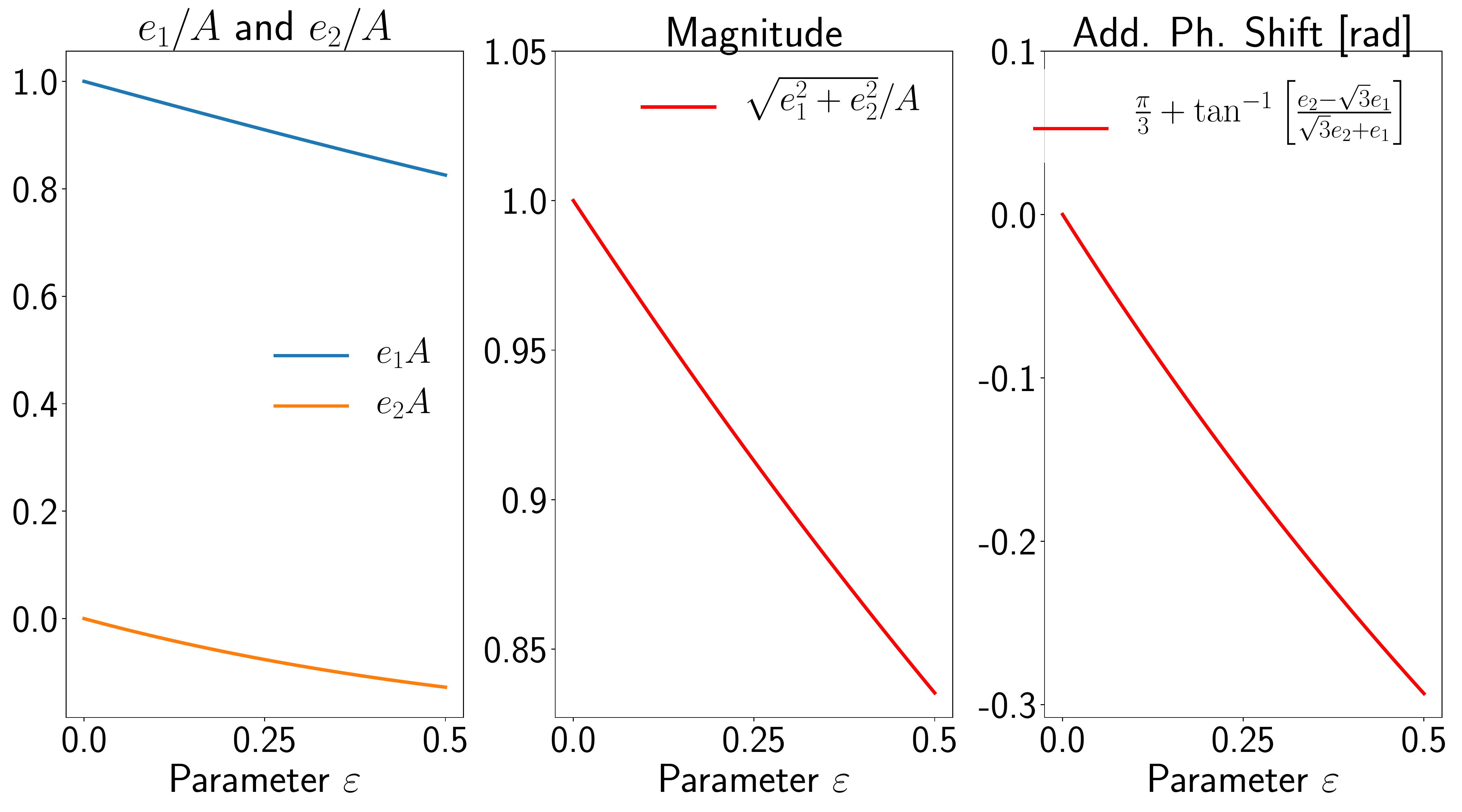}
\caption{Parameters $e_1$ and $e_2$ as a function of $\varepsilon$, Magnitude $\sqrt{e_1^2+e_2^2}$ and Additional Phase Shift $\frac{\pi}{3} + \arctan\left( \frac{e_2 - \sqrt{3}e_1}{\sqrt{3}e_2 + e_1} \right)$ of the Limiting Sinusoid}
\label{fig:Price_Memory_Variable_Limiting_Sinusoid}
\end{figure}

\begin{figure}[hbtp]
\centering
\includegraphics[scale=0.19]{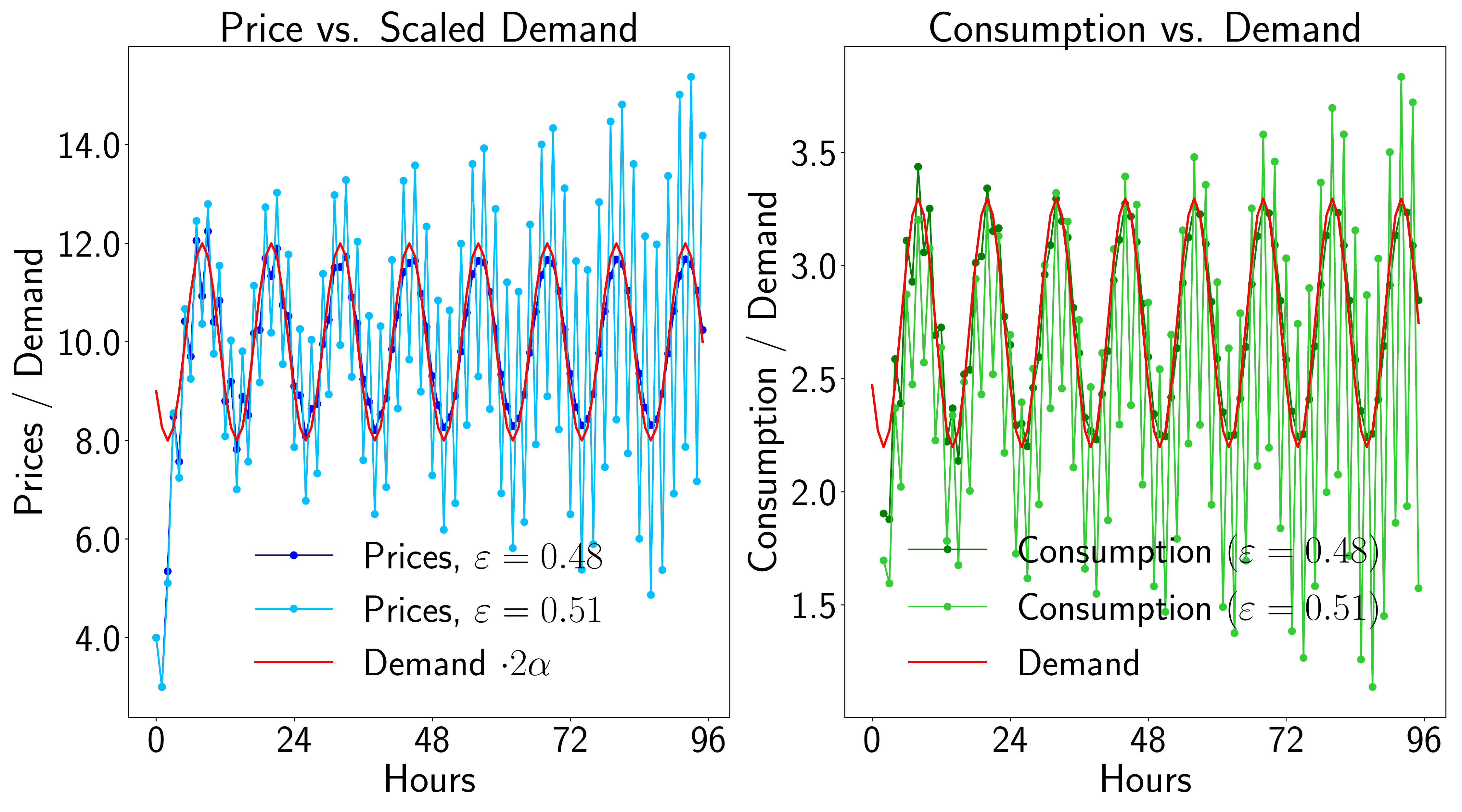}
\caption{Evolution of Prices and Consumption, Variable Demand, $\beta=4$, $\alpha = 1.92$ (Stable) or $\alpha = 2.04$ (Unstable). Initial Conditions $\lambda_0 = 2, \lambda_1 = 3$}
\label{fig:Price_Recurrence_Variable_Dem_Stable_Unstable}
\end{figure}

Figure \ref{fig:Price_Recurrence_Variable_Dem_Stable_Unstable} shows the evolution of prices \eqref{eq:soln_rec_variable} for $\varepsilon = 0.48 < 0.5$, which is a stable system, and for $\varepsilon = 0.51 > 0.5$, which results in an unstable system. It is clearly seen that the stable price trajectory with $\varepsilon = 0.48$ converges to a limiting sinusoid with the same mean, but a smaller magnitude ($\approx 0.84 \cdot A$) than the scaled demand $2\alpha d_k$, see Figure \ref{fig:Price_Memory_Variable_Limiting_Sinusoid}.

\subsection{Stability Under Consumption Model with Price and Consumption Memory}
Here, we consider \eqref{eq:cons_opt_flattening} as a consumption model. As in the previous case, let $c(x) := \alpha x^2 \therefore \dot{c}(x) = 2\alpha x$, and $V(x) := \gamma x^2 \therefore \dot{V}^{-1}(x) = \frac{1}{2\gamma}x = \tilde{V}x,~\alpha, \gamma \neq 0$. Let $\tilde{\varepsilon}:=\alpha/\gamma$. Then the following coupled system establishes a dynamical system for the evolution of prices and consumption:
\begin{subequations}
\begin{align}
u_k &= \frac{d_k + \tilde{V}\left(\lambda_{k-1} -\lambda_k + 2\rho(d_k - d_{k-1} + u_{k-1}) \right)}{1 + 2\rho \tilde{V}} \label{eq:recurr_price_cons_memory_u}\\
\lambda_{k+1} &= \dot{c}(\hat{u}_{k+1}) = 2\alpha u_k \label{eq:recurr_price_cons_memory_lambda}
\end{align}
\end{subequations}

Eliminating $u_k$ and $u_{k-1}$ from \eqref{eq:recurr_price_cons_memory_u} and \eqref{eq:recurr_price_cons_memory_lambda}, we obtain the following difference equation for the prices:
\begin{equation}\label{eq:recurr_prices_only_p_c_memory}
\lambda_{k+1} = \frac{\rho-\alpha}{\gamma+\rho}\lambda_k + \frac{\alpha}{\gamma+\rho}\lambda_{k-1} + 2\alpha d_k - \frac{2\alpha\rho}{\gamma+\rho}d_{k-1}
\end{equation} 
Note that we recover \eqref{eq:price_dynamics_memory_linear} for $\rho = 0$ and $\gamma = \beta$.

\begin{theorem}\label{thm:rho_stability_conditions}
Independently of $\rho$, \eqref{eq:recurr_prices_only_p_c_memory} is stable for $0 \leq \alpha < \gamma/2$. For $\alpha \geq \gamma/2$, \eqref{eq:recurr_prices_only_p_c_memory} is stable for $\rho > \alpha - \gamma/2 > 0$, or equivalently, $\tilde{\varepsilon} < 1/2 + \rho/\gamma$.
\end{theorem}

Theorem \ref{thm:rho_stability_conditions} states that an appropriate choice of $\rho$ can turn unstable dynamics from the consumption model with price memory only (namely for $\alpha \geq \gamma/2$) into a stable system.

\subsubsection{Constant Demand}\label{sec:price_cons_model_constant_demand}
For the special case of constant demand $d$, we obtain the following difference equation:
\begin{equation}\label{eq:recurr_const_demand_p_c_memory}
\lambda_{k+1} = \frac{\rho - \alpha}{\gamma + \rho}\lambda_k + \frac{\alpha}{\gamma + \rho}\lambda_{k-1} + \frac{2\alpha\gamma}{\gamma + \rho}d.
\end{equation}
Assuming that the price converges, the fixed points $\lambda^\ast$ and $u^\ast$ obtained from \eqref{eq:recurr_price_cons_memory_u} and \eqref{eq:recurr_price_cons_memory_lambda} are identical to the consumption model with price memory only \eqref{eq:fixed_point_lambda_and_u}.

\begin{figure}[hbtp]
\centering
\includegraphics[scale=0.19]{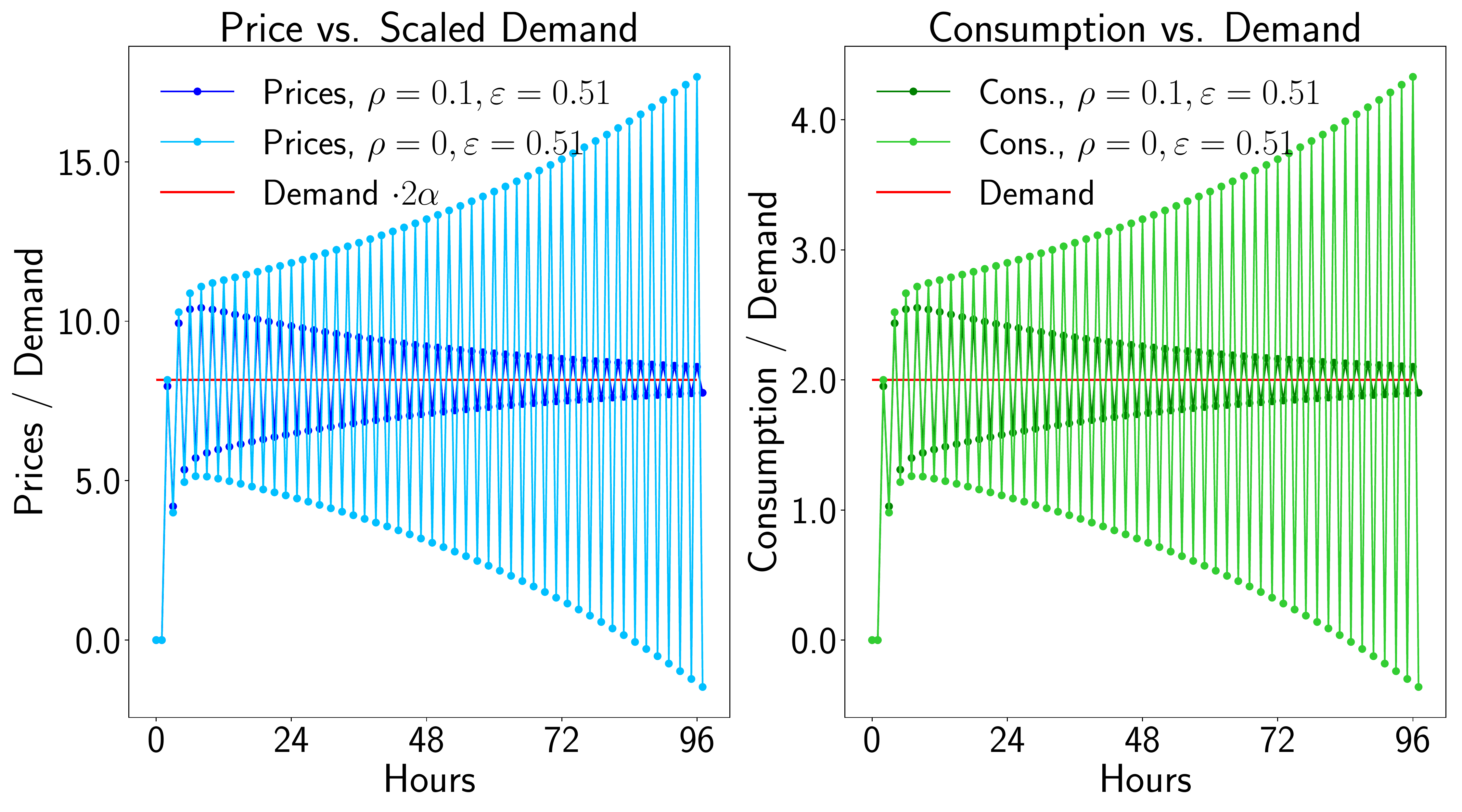}
\caption{Evolution of Prices and Consumption, Constant Demand, $\gamma=4$, $\alpha = 2.04$, and $\rho=0$ (Unstable) or $\rho = 0.1$ (Stable). Initial Conditions $\lambda_0 = \lambda_1 = 0$}
\label{fig:Price_Recurrence_Const_Dem_Stabilization}
\end{figure}

Figure \ref{fig:Price_Recurrence_Const_Dem_Stabilization} shows the evolution of prices and the consumption for $\alpha = 2.04, \gamma = 4.0 \Rightarrow \tilde{\varepsilon} = 0.51 > 0.5$. The instability of the price and consumption evolution for $\rho=0$ (which corresponds to the consumption model with pure price memory) is consistent with the findings of Section \ref{sec:Consumption_Price_Memory}. However, any choice of $\rho > \alpha - \gamma/2 = 0.04$ (Theorem \ref{thm:rho_stability_conditions}) stabilizes the system (e.g. we chose $\rho = 0.1$), as is shown in the figure.

\begin{figure}[hbtp]
\centering
\vspace{-0.1cm}
\includegraphics[scale=0.19]{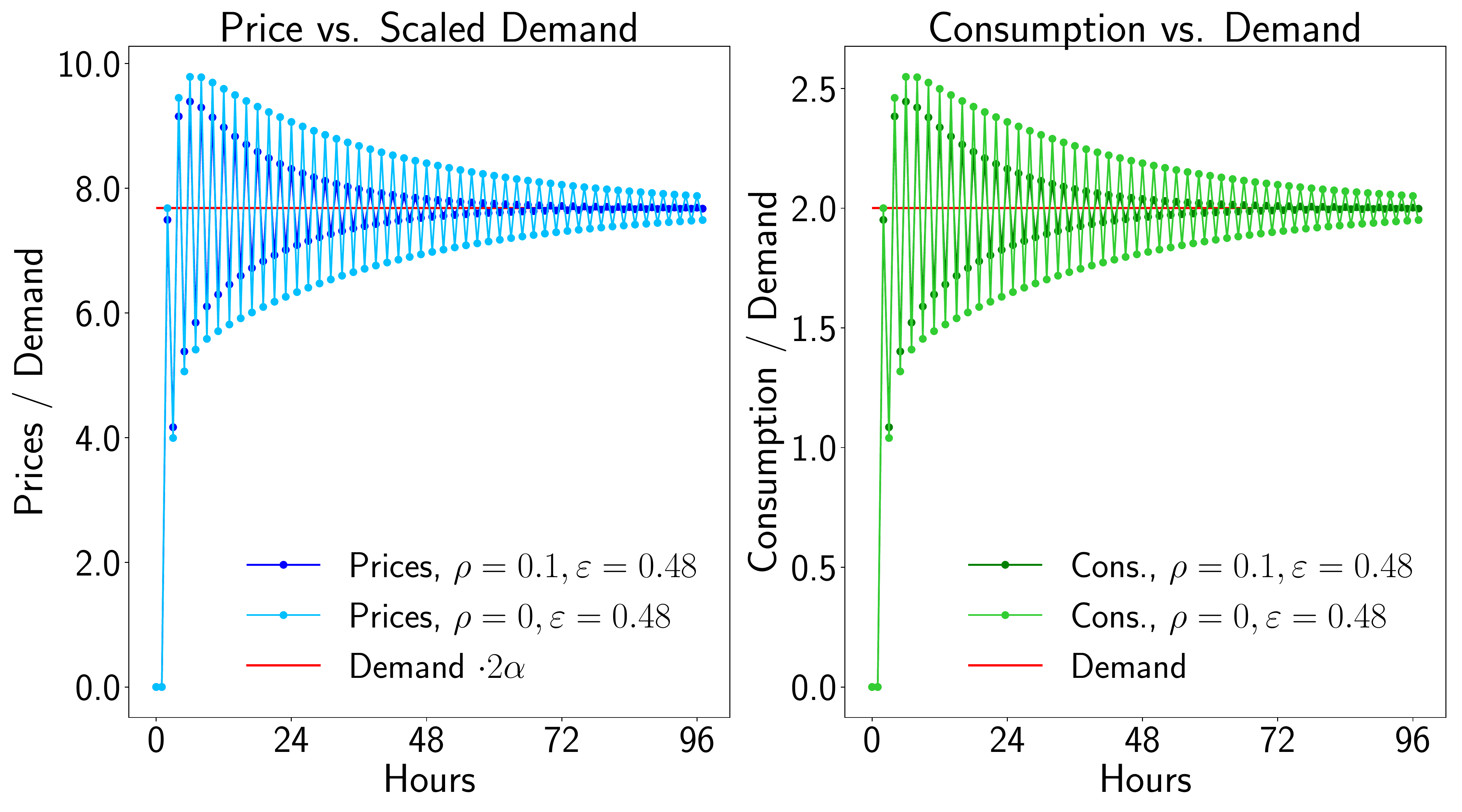}
\caption{Evolution of Prices and Consumption, Constant Demand, $\gamma=4$, $\alpha = 2.04$, and $\rho = 0$ (No Flattening) or $\rho = 0.1$ (With Flattening). Initial Conditions $\lambda_0 = \lambda_1 = 0$}
\label{fig:Price_Recurrence_Const_Dem_Flattening}
\vspace{-0.3cm}
\end{figure}

With an appropriately chosen weight $\rho$, we further see a flattening of the price and consumption evolution, and a faster convergence towards the equilibria $u^\ast$ and $\lambda^\ast$, as is shown in Figure \ref{fig:Price_Recurrence_Const_Dem_Flattening}.

\subsubsection{Variable Demand}\label{sec:price_cons_model_variable_demand}
The same qualitative findings as under the constant demand approximation are found in this case, that is, for increasing values of $\rho=0.1$, there is a flattening of the price and consumption evolution as well as a faster convergence towards the limiting sinusoid.

%% file: Sections/Conclusion.tex
\section{Conclusion}
\label{sec:Conclusion}
We proposed realistic electricity consumption models with price and consumption memory by using dynamic programming on a suitably defined inventory problem of the elastic demand over the consumption horizon of a user. These dynamic consumption models were then analyzed in a closed-loop real-time market setting with representative agents for supply and demand under the assumption of quadratic generation costs and backlog disutilities, and the resulting electricity price and demand evolution were investigated for stability and dynamic properties. For the consumption model with price memory only, we showed that prices remain stable if the marginal generation cost is less than half of the marginal backlog disutility of the consumers. In addition, by penalizing deviations from required demand in the consumption model with price and consumption memory, this ratio could be increased beyond $1/2$, with the additional effect of faster convergence towards the equilibrium price.

Our analyses could be used as a preliminary guidepost for the design of real-time pricing mechanisms to diminish the justifiable reluctance of ratepayers towards adoption of real-time pricing. For a more confident analysis, however, Assumptions \ref{as:cost_function_convex}-\ref{as:c_linear_function}, the Representative Agent Theorem, and the absence of transmission, capacity, and congestion constraints would need to be removed. The price dynamics in a resulting network of users is subject to future research.

We are currently working on a contractual formulation between load serving entities and end-users for real-time pricing mechanisms in a Demand Response setting with the goal to find optimal bidding strategies into the wholesale electricity market, which is a direct continuation of the work described in this paper.

%% file: Sections/Appendix.tex
%
%


\section*{Appendix}
\label{sec:Appendix}

\subsection*{Proof of Theorem \ref{thm:optimal_consumption_bellmann}}
For $k=1$, the terminal constraint $x_n=0$ forces the optimal consumption to be $u_{n-1}^\ast = d_{n-1} - x_{n-1}$. The $k=1$ cost-to-go is then obtained as
\begin{equation*}
J_{n-1}^\ast = \underset{u_{n-1}}{\text{min}}\left[ \lambda_{n-1} u_{n-1} + p(x_n) \right] = \lambda_{n-1}(d_{n-1} - x_{n-1}),
\end{equation*}
which equals \eqref{eq:opt_costtogo_bmann} for $k=1$.
For $k = n-2$, we need to solve
\begin{equation*}
\begin{aligned}
J_{n-2}^\ast =~ & \underset{u_{n-2}} {\text{min}} & &  \lambda_{n-2} u_{n-2} + p(x_{n-1}) + \mathbb{E}_{\lambda_{n-1}}\left[ J_{n-1}^\ast \right]\\
&~ \text{s.t.} & & x_{n-1} = x_{n-2} + u_{n-2} - d_{n-2}
\end{aligned}
\end{equation*}
With $J_{n-1}^\ast$ and Assumption \ref{as:expected_price}, the first order optimality condition is
\begin{align}
0 &= \lambda_{n-2} + \frac{dp(x_{n-1})}{dx_{n-1}}\frac{dx_{n-1}}{du_{n-2}} - \mathbb{E}[\lambda_{n-1}]\nonumber\\
x_{n-1} &= \dot{p}^{-1}\left(\underbrace{\mathbb{E}\left[ \lambda_{n-1} \right] - \lambda_{n-2}}_{=0} \right) = x_{n-2}+ u_{n-2} - d_{n-2}\nonumber\\
u_{n-2}^\ast &= \dot{p}^{-1}(-\lambda_{n-2}+\lambda_{n-3}) + d_{n-2}\label{eq:opt_cons_nminus2}
\end{align}
\eqref{eq:opt_cons_nminus2} is identical to \eqref{eq:opt_consumption_bmann} for $k=2$. The $n-2$ cost-to-go is found by plugging \eqref{eq:opt_cons_nminus2} back into the cost function, and is identical to \eqref{eq:opt_costtogo_bmann} for $k=2$. With $k=2$ as the base case, we proceed to the induction step at $n-k-1$:
Solving for the optimal consumption $u_{n-k-1}^\ast$ in
\begin{equation*}
\begin{aligned}
J_{n-k-1}^\ast =~ & \underset{u_{n-k-1}} {\text{min}} & &  \lambda_{n-k-1} u_{n-k-1} + p(x_{n-k}) + \mathbb{E} J_{n-k}^\ast\\
&~ \hspace{0.12cm} \text{s.t.} & & x_{n-k} = x_{n-k-1} + u_{n-k-1} - d_{n-k-1}
\end{aligned}
\end{equation*}
yields
\begin{align}
0 &= \lambda_{n-k-1} + \frac{d(p(x_{n-k})+J_{n-k})}{dx_{n-k}}\frac{dx_{n-k}}{du_{n-k-1}}\nonumber\\
x_{n-k} &= \dot{p}^{-1}\left( (\mathbb{E}[\lambda_{n-k}] - \lambda_{n-k-1})/k \right) \nonumber\\
&= x_{n-k-1} + u_{n-k-1} - d_{n-k-1}\nonumber\\
u_{n-k-1}^\ast &= d_{n-k-1} - \dot{p}^{-1}\left((\lambda_{n-k} - \lambda_{n-k-1})/k\right)\label{eq:opt_cons_nminuskminus1}
\end{align}
Plugging \eqref{eq:opt_cons_nminuskminus1} into the cost function yields, after simplifications and Assumption \ref{as:expected_price},
\begin{align}
J_{n-k-1}^\ast &= \lambda_{n-k-1} d_{n-k-1} \nonumber\\
&\quad +k\cdot p \left( x_{n-k-1} - \dot{p}^{-1}((\lambda_{n-k-1} - \lambda_{n-k-2})/k \right)\nonumber\\
&\quad +\mathbb{E}\left[\lambda_{n-k}\right]\left( \sum_{i=n-k}^{n-1} d_i - x_{n-k-1} \right)\nonumber\\
J_{n-k-1}^\ast &= k\cdot p \left( x_{n-k-1} - \dot{p}^{-1}((\lambda_{n-k-1} - \lambda_{n-k-2})/k \right)\nonumber\\
&\quad +\lambda_{n-k-1}\left( \sum_{i=n-k-1}^{n-1} d_i - x_{n-k-1} \right)\label{eq:opt_cost_nminuskminus1}
\end{align}
\eqref{eq:opt_cons_nminuskminus1} and \eqref{eq:opt_cost_nminuskminus1} are \eqref{eq:opt_consumption_bmann} and \eqref{eq:opt_costtogo_bmann} evaluated at $n-k-1$, respectively, which completes the induction step and the proof.
 
\subsection*{Proof of Theorem \ref{thm:stability_price_memory}}
The characteristic equation of the homogeneous portion is 
\begin{equation*}
x^2 + \varepsilon x - \varepsilon = 0
\end{equation*}
and the roots are 
\begin{equation}\label{eq:roots_linear_price}
x_{1,2} = \frac{-\varepsilon}{2} \pm \frac{\sqrt{\varepsilon(\varepsilon+4)}}{2}
\end{equation}

and so we see that the roots are always real valued and of magnitude less than 1 if and only if $0 \leq \varepsilon < 1/2$, which is the condition for stability. 
 
\subsection*{Proof of Theorem \ref{thm:price_dyn_price_memory}}
The homogeneous solution to \eqref{eq:price_dynamics_memory_linear} can be found by solving for the roots of its characteristic equation
\begin{align*}
\lambda^2 + \frac{\alpha}{\beta} \lambda - \frac{\alpha}{\beta} = 0
\end{align*}
Combining the homoegeneous solution with the particular solution $\lambda_p = 2\alpha d$ yields \eqref{eq:price_dynamics_solution_price_memory}.
 
\subsection*{Proof of Theorem \ref{thm:price_dynamics_price_cons_memory}}
A particular solution to the price dynamics can be found with the guess
\begin{equation}\label{eq:particular_solution_periodic_demand}
b_k = e_0 + e_1 \sin\left(\frac{(k-5)\pi}{6}\right) + e_2\cos\left(\frac{(k-5)\pi}{6}\right).
\end{equation}
\eqref{eq:particular_solution_periodic_demand} needs to solve \eqref{eq:price_dynamics_memory_linear}, i.e. 
\begin{equation*}
b_k + \varepsilon b_{k-1} -\varepsilon b_{k-2} = \mu + A\sin\left(\frac{(k-5)\pi}{6}\right)
\end{equation*}

Using the trigonometric identities
\begin{subequations}
\begin{align}
\sin(\alpha + \beta) = \sin\alpha \cos\beta + \cos\alpha \sin\beta \label{eq:trigon_1}\\ 
\cos(\alpha + \beta) = \cos\alpha \cos\beta - \sin\alpha \sin\beta \label{eq:trigon_2}
\end{align}
\end{subequations}
the coefficients $e_0, e_1, e_2$ of the particular solution can be determined as
\begin{align}
e_0 &= \mu \\
e_1 &= \frac{1 + \varepsilon(\sqrt{3}-1)/2}{1 + (\sqrt{3}-1)\varepsilon + (2 - \sqrt{3})\varepsilon^2}A \\
e_2 &= \frac{\varepsilon(1-\sqrt{3})/2}{1 + (\sqrt{3}-1)\varepsilon + (2 - \sqrt{3})\varepsilon^2}A 
\end{align}

\subsection*{Proof of Theorem \ref{thm:price_traj_limit}}
For $\varepsilon < 1/2$, we have that $|x_1|, |x_2| < 1$, and so if $k$ goes to infinity, the first two terms in \eqref{eq:soln_rec_variable} go to zero. Using the following identity \cite{Kythe:2014aa}
\begin{equation}
\begin{aligned}
A\cos(\omega t + \alpha) + B\sin(\omega t + \alpha) = \sqrt{A^2 + B^2}~\times\nonumber\\
\cos\left( \omega t + \arctan\left( \frac{A\sin\alpha - B\cos\alpha}{A\cos\alpha + B\sin\alpha} \right) \right)
\end{aligned}
\end{equation}
on \eqref{eq:soln_rec_variable}, together with $\sin(x+\pi/2) = \cos(x)$ and $\cos(x-\pi/2) = \sin(x)$, and noting that $e_2 < 0$, as is shown in Figure \ref{fig:Price_Memory_Variable_Limiting_Sinusoid}, \eqref{eq:price_traj_limit} is readily obtained.

\subsection*{Proof of Theorem \ref{thm:rho_stability_conditions}}
The stability and the dynamic behavior of \eqref{eq:recurr_prices_only_p_c_memory} can be checked by analyzing the characteristic equation, i.e.
\begin{equation*}
x^2 - \frac{\rho - \alpha}{\gamma + \rho}x - \frac{\alpha}{\gamma + \rho} = 0
\end{equation*}

The roots to the characteristic equation are
\begin{equation}\label{eq:roots_char_equation_p_c_memory}
x_{1,2} = \frac{\rho - \alpha \pm \sqrt{(\rho-\alpha)^2 + 4\alpha(\gamma+\rho)}}{2(\gamma+\rho)}.
\end{equation}

Stability is guaranteed if and only if the roots $x_1, x_2$ to \eqref{eq:roots_char_equation_p_c_memory} have magnitude less than 1. First we analyze $x_2$. The values of $\rho$ for which $x_2 <1$ are determined:
\begin{align}
(\rho-\alpha) - 2(\gamma+\rho) &< -\sqrt{(\rho-\alpha)^2 + 4\alpha(\gamma+\rho)} \label{eq:x2_root_conditions}\\
((\rho-\alpha) - 2(\gamma+\rho))^2 &> (\rho-\alpha)^2 + 4\alpha(\gamma+\rho) \nonumber\\
\alpha + \gamma &> \alpha \nonumber
\end{align}
Note that both sides of \eqref{eq:x2_root_conditions} are negative, which forces an inequality switch. The last inequality is valid for all $\rho$, and so $x_2$ is always $< 1$ for all values of $\rho$.

Second, we find the value of $\rho$ for which $x_1 = -1$:
\begin{align}
x_1 &= \frac{\rho - \alpha - \sqrt{(\rho-\alpha)^2 + 4\alpha(\gamma+\rho)}}{2(\gamma+\rho)} = -1 \label{eq:left_root_minus1}\\
\rho &= \alpha - \gamma/2 \nonumber
\end{align}

We have seen that in this case, for $\rho = 0$, the system is stable (see Section \ref{sec:Constant_Demand}). Also, note that the ``left" root $x_1$ is always smaller than the ``right" root, i.e. $x_1 \leq x_2$, and so $x_1$ lower bounds $x_2$. Take the derivative of $x_1$ w.r.t. $\rho$:
\begin{equation*}
\begin{aligned}
\frac{dx_1}{d\rho} = &\frac{1}{2}\left(\gamma+\rho\right)^{-2}\left((\gamma+\rho)\underbrace{\left(1 - \frac{\rho+\alpha}{\sqrt{(\rho+\alpha)^2 + 4\alpha\gamma}} \right)}_{\geq 0} + \right.\\
&\left. \underbrace{\sqrt{(\rho-\alpha)^2 + 4\alpha(\gamma+\rho)} - (\rho-\alpha)}_{\geq 0}\right) \geq 0.
\end{aligned}
\end{equation*}
Thus, the left root is monotonely nondecreasing in $\rho$, and we can take the limit of \eqref{eq:left_root_minus1}:
\begin{equation*}
\lim_{\rho\rightarrow\infty} x_1 = (1 - \sqrt{1})/2 = 0.
\end{equation*}
Thus, $x_1 = -1$ for $\rho=\alpha - \gamma/2$, and for increasing values of $\rho$, $x_1$ approaches the value of zero. As $x_1$ lower bounds $x_2$ and $-1 < x_1 \leq x_2 < 1$ for $\rho \geq \alpha - \gamma/2$, stability is found if and only if $\rho \geq \alpha - \gamma/2$. Note that if $\alpha < \gamma/2$, the system is always stable due to $\rho > 0$. 

%% file: Market_Stability.bbl
\begin{thebibliography}{10}
\providecommand{\url}[1]{#1}
\csname url@samestyle\endcsname
\providecommand{\newblock}{\relax}
\providecommand{\bibinfo}[2]{#2}
\providecommand{\BIBentrySTDinterwordspacing}{\spaceskip=0pt\relax}
\providecommand{\BIBentryALTinterwordstretchfactor}{4}
\providecommand{\BIBentryALTinterwordspacing}{\spaceskip=\fontdimen2\font plus
\BIBentryALTinterwordstretchfactor\fontdimen3\font minus
  \fontdimen4\font\relax}
\providecommand{\BIBforeignlanguage}[2]{{%
\expandafter\ifx\csname l@#1\endcsname\relax
\typeout{** WARNING: IEEEtran.bst: No hyphenation pattern has been}%
\typeout{** loaded for the language `#1'. Using the pattern for}%
\typeout{** the default language instead.}%
\else
\language=\csname l@#1\endcsname
\fi
#2}}
\providecommand{\BIBdecl}{\relax}
\BIBdecl

\bibitem{Borenstein:2005aa}
S.~Borenstein, ``{The Long-Run Efficiency of Real-Time Electricity Pricing},''
  \emph{The Energy Journal}, 2005.

\bibitem{Allcott:2011aa}
H.~Allcott, ``{Rethinking Real Time Electricity Pricing},'' \emph{Resource and
  Energy Economics}, vol.~33, no.~4, pp. 820--842, 2011.

\bibitem{Borenstein:2002aa}
S.~Borenstein, M.~Jaske, and A.~Rosenfeld, ``{Dynamic Pricing, Advanced
  Metering, and Demand Response in Electricity Markets},'' \emph{University of
  California Energy Institute, Center for the Study of Energy Markets}, 2002.

\bibitem{Borenstein:2005ab}
S.~Borenstein and S.~P. Holland, ``{On the Efficiency of Competitive
  Electricity Markets with Time-Invariant Retail Prices},'' \emph{Rand Journal
  of Economics}, vol.~36, no.~3, pp. 469--493, 2005.

\bibitem{Mohsenian-Rad:2010aa}
A.-H. Mohsenian-Rad, V.~W.~S. Wong, J.~Jatskevich, R.~Schober, and
  A.~Leon-Garcia, ``{Autonomous Demand-Side Management Based on Game-Theoretic
  Energy Consumption Scheduling},'' \emph{IEEE Transactions on Smart Grid},
  2010.

\bibitem{Conejo:2010aa}
A.~J. Conejo, J.~M. Morales, and L.~Baringo, ``{Real-Time Demand Response
  Model},'' \emph{IEEE Transactions on Smart Grid}, vol.~1, no.~3, 2010.

\bibitem{Chen:2010aa}
Z.~Chen, L.~Wu, and Y.~Fu, ``{Real-Time Price-Based Demand Response Management
  for Residential Appliances via Stochastic Optimization and Robust
  Optimization},'' \emph{IEEE Transactions on Smart Grid}, vol.~3, no.~4, 2010.

\bibitem{Faghih:2011aa}
A.~Faghih, M.~Roozbehani, and M.~Dahleh, ``{Optimal Utilization of Storage and
  the Induced Price Elasticity of Demand in the Presence of Ramp
  Constraints},'' \emph{Conference on Decision and Control}, 2011.

\bibitem{Bitar:2011aa}
E.~Bitar, R.~Rajagopal, P.~Khargonekar, and K.~Poolla, ``{The Role of Colocated
  Storage for WPP in Conventional Electricity Markets},'' \emph{American
  Control Conference}, 2011.

\bibitem{Huang:2015aa}
Q.~Huang, M.~Roozbehani, and M.~Dahleh, ``{Efficiency-Risk Tradeoffs in
  Electricity Markets with Dynamic Demand Response},'' \emph{IEEE Transactions
  on Smart Grid}, vol.~6, no.~1, 2015.

\bibitem{Roozbehani:2011aa}
M.~Roozbehani, ``{Analysis of Competitive Electricity Markets under a New Model
  of Real-Time Retail Pricing},'' \emph{8th International Conference on the
  European Energy Market (EEM)}, 2011.

\bibitem{Alvarado:1999aa}
F.~Alvarado, ``{The Stability of Power System Markets},'' \emph{IEEE
  Transactions on Power Systems}, vol.~14, no.~2, 1999.

\bibitem{Roozbehani:2014aa}
M.~Roozbehani, D.~Materassi, M.~Ohannessian, and M.~Dahleh, ``{Robust and
  Optimal Consumption Policies for Deadline-Constrained Deferrable Loads},''
  \emph{IEEE Transactions on Smart Grid}, 2014.

\bibitem{Li:2011aa}
N.~Li, L.~Chen, and S.~Low, ``{Optimal Demand Response Based on Utility
  Maximization in Power Networks},'' \emph{IEEE Power and Energy Society
  General Meeting}, 2011.

\bibitem{Wu:1996aa}
F.~Wu, P.~Varaiya, P.~Spiller, and S.~Oren, ``{Folk Theorems on Transmission
  Access: Proofs and Counterexamples},'' \emph{Journal of Regulatory
  Economics}, pp. 1015--1023, 1996.

\bibitem{Gomez-Exposito:2016aa}
A.~Gomez-Exposito, A.~J. Conejo, and C.~Canizares, \emph{Electric Energy
  Systems: Analysis and Operation}.\hskip 1em plus 0.5em minus 0.4em\relax CRC
  Press, 2016.

\bibitem{Roozbehani:2012aa}
M.~Roozbehani, M.~Dahleh, and S.~K. Mitter, ``{Volatiliy of Power Grids under
  Real-Time Pricing},'' \emph{IEEE Transactions on Power Systems}, 2012.

\bibitem{Canova:2007aa}
F.~Canova, \emph{Methods for Applied Macroeconomic Research}.\hskip 1em plus
  0.5em minus 0.4em\relax Princeton University Press, 2007.

\bibitem{Zipkin:2000aa}
P.~H. Zipkin, \emph{{Foundations of Inventory Management}}.\hskip 1em plus
  0.5em minus 0.4em\relax McGraw-Hill/Irwin, 2000.

\bibitem{Zhou:2016aa}
D.~Zhou, M.~Balandat, and C.~Tomlin, ``{Residential Demand Response Targeting
  Using Machine Learning with Observational Data},'' \emph{55th IEEE Conference
  on Decision and Control}, 2016.

\bibitem{Kythe:2014aa}
P.~K. Kythe, \emph{{Sinusoids: Theory and Applications}}.\hskip 1em plus 0.5em
  minus 0.4em\relax CRC Press, 2014.

\end{thebibliography}
